\documentclass[journal]{IEEEtran}
\IEEEoverridecommandlockouts
\usepackage{cite}
\usepackage{easybmat}
\usepackage{stfloats}
\usepackage{amsmath,amssymb,amsfonts}
\usepackage{graphicx}
\usepackage{textcomp}
\usepackage{epstopdf}
\usepackage[printonlyused,withpage]{acronym}
\usepackage[dvipsnames]{xcolor}
\usepackage{caption}
\usepackage{subcaption}
\usepackage{algorithm}
\usepackage{algorithmic}
\usepackage{multirow}
\usepackage{tabularx}

\usepackage{setspace}

\DeclareFontFamily{U}{mathx}{\hyphenchar\font45}
\DeclareFontShape{U}{mathx}{m}{n}{
      <5> <6> <7> <8> <9> <10>
      <10.95> <12> <14.4> <17.28> <20.74> <24.88>
      mathx10
      }{}
\DeclareSymbolFont{mathx}{U}{mathx}{m}{n}
\DeclareFontSubstitution{U}{mathx}{m}{n}
\DeclareMathAccent{\widecheck}{0}{mathx}{"71}

\usepackage[letterpaper, left=0.7in, right=0.7in, bottom=1in, top=0.7in]{geometry}

\newtheorem{thm}{Theorem}
\newtheorem{prop}{Proposition}
\newtheorem{rmk}{Remark}

\newcolumntype{Y}{>{\centering\arraybackslash}X}

\newenvironment{proof}{ \paragraph*{\hspace{-1em}Proof}}{\hfill$\square$}

\renewcommand{\vec}[1]{\ensuremath{\boldsymbol{#1}}}
\newcommand{\bs}[1]{\ensuremath{\boldsymbol{#1}}}
\newcommand{\be}{\begin{equation}}
\newcommand{\ee}{\end{equation}}
\newcommand{\ba}{\begin{array}}
\newcommand{\ea}{\end{array}}
\newcommand{\bea}{\begin{eqnarray}}
\newcommand{\eea}{\end{eqnarray}}
\newcommand{\A}{\vec{A}}

\newcommand{\Btil}{\widetilde{\boldsymbol{B}}}

\newcommand{\W}{\vec{W}}

\renewcommand{\H}{\vec{H}}
\newcommand{\mrm}[1]{\ensuremath{\mathrm{#1}}}

\newcommand{\diag}{\mathrm{diag}} 

\newcommand{\tA}{\widetilde{\A}}

\newcommand{\comment}[1]{}

\newcommand{\MP}{M_\mrm{P}}
\newcommand{\TP}{T_\mrm{P}}

\acrodef{AWGN}{additive white Gaussian noise}
\acrodef{MIMO}{mutiple-input multiple-output}
\acrodef{RIS}{reconfigurable intelligent surface}
\acrodef{BS}{base station}
\acrodef{5G}{5th generation of mobile broadband communications}
\acrodef{MU-MIMO}{multi-user \ac{MIMO}}
\acrodef{UE}{user equipment}
\acrodef{CPU}{central processing unit}
\acrodef{BB}{baseband}
\acrodef{BBU}{\ac{BB} unit}
\acrodef{UL}{uplink}
\acrodef{DL}{downlink}
\acrodef{SNR}{signal-to-noise ratio}
\acrodef{MRC}{maximum ratio combining}

\begin{document}

\title{Trade-Offs in Decentralized Gigantic MIMO with Hard-Boundary Constraints}
\author{Juan Vidal Alegr\'ia,
        Joao Vieira, and Ove Edfors

\thanks{This work was supported by "SSF Large Intelligent Surfaces - Architecture and Hardware" Project CHI19-0001, and by the NextG2Com competence center (VINNOVA Grant 2023-00541).}
\thanks{J. Vidal Alegr\'{i}a, and O. Edfors are with the Department of Electrical and Information Technology, Lund University, 221 00 Lund, Sweden. (email: \{juan.vidal\_alegria@eit.lth.se; ove.edfors@eit.lth.se)

J. Vieira is with Ericsson Research, 223 62 Lund, Sweden (e-mail: joao.vieira@ericsson.com).}}

\maketitle

\begin{abstract}
To maintain the antenna apertures offered by 5G massive MIMO systems operating at the sub-6GHz band, known as FR1, 6G base stations (BSs) using the upper-mid band, FR3, should increase the number of antennas by a factor 4-8, giving rise to gigantic MIMO. This poses challenges in terms of processing complexity and interconnection bandwidth. The WAX framework, previously introduced for exploring trade-offs in decentralized architectures, may offer the flexibility needed to tackle these challenges. However, no results have been established on the applicability of this framework in the presence of hard-boundary constraints. The current work explores gigantic MIMO implementations based on a novel adaptation of the WAX framework, where the decentralized processing is performed by non-cooperating hardware modules. These modules may be implemented through state-of-the-art massive MIMO baseband units (BBUs). The results show the potential of the proposed framework towards exploiting trade-offs between complexity and performance in practical gigantic MIMO implementations.
\end{abstract}

\begin{IEEEkeywords}
WAX decomposition, distributed MIMO, massive MIMO, gigantic MIMO, decentralized processing.
\end{IEEEkeywords}

\section{Introduction}
\label{section:intro}
\IEEEPARstart{M}{assive} \ac{MIMO} \cite{marzetta} has become a key enabling technology towards the current \ac{5G} \cite{mMIMO_reality}. Extending \ac{MU-MIMO} by increasing the number of antennas at the \ac{BS}, massive \ac{MIMO} achieves impressive gains in terms of spectral efficiency, while relaxing the implementation requirements due to its averaging effects on general imperfections. 

As we transition towards the upcoming 6G, new frequency bands are being explored to accommodate the increased data rate demands. The upper-mid band (7-24~GHz), also known as FR3, is a strong candidate towards 6G since it provides a good compromise between available spectrum, reasonably low propagation loss, and ease of hardware implementation \cite{g_mimo}. However, to maintain approximately the same antenna aperture with respect to the sub-6GHz band (FR1) employed in \ac{5G} massive \ac{MIMO}, in other words, to compensate for the increase in propagation loss as we go up in frequency, the upper-mid band requires increasing the number of antennas by a factor $4$-$8$. This leads to the new paradigm of gigantic \ac{MIMO}, which corresponds to a further scaling of massive \ac{MIMO} to enable operability in the upper-mid band \cite{g_mimo}.


The research literature directly related to the implementation aspects of gigantic \ac{MIMO}, as proposed in \cite{g_mimo}, is not very extensive. In \cite{anal_g_mimo}, analog processing is explored in order to facilitate the scalability of gigantic \ac{MIMO}. On the other hand, \cite{g_mimo_caire} explores the potential of \acp{RIS} as a smart antenna technology to enable the implementation of gigantic \ac{MIMO} transceivers. Instead, we propose an alternative approach consisting of distributing the \ac{BB} processing into a set of decentralized modules by exploiting a general framework considered in \cite{wax_journal,A_struct_journal,cell-free_wax}, hereby referred to as the WAX framework. In \cite{nir_wax}, the same framework was considered under the term modular hybrid receive processing, where the optimization of the decentralized filters was performed using learning-based approaches. We however focus on the original formulation with the aim at exploiting its associated trade-offs to design suitable decentralized processing strategies for gigantic \ac{MIMO}.

The current work explores how to exploit the WAX framework from \cite{wax_journal,A_struct_journal,cell-free_wax} to enable the integration of the processing performed at decentralized modules, as well as of that performed at a \ac{CPU}, into state-of-the-art massive \ac{MIMO} \acp{BBU}. To the best of our knowledge, this problem has not been analyzed in previous literature. We believe that the proposed work may facilitate scalability, as well as backward-compatibility, in the transition from massive \ac{MIMO} to gigantic \ac{MIMO}. Moreover, the considered framework may be adapted to perform decentralized processing in the analog domain instead of in the \ac{BB} domain, as considered in \cite{anal_g_mimo,g_mimo_caire}. Said adaptation may consist of a combination of the results presented in this work with those presented in \cite{asilomar24}, which considers a modified WAX framework including analog processing constraints.

\section{System model}\label{section:model}
We consider a narrowband \ac{UL} \ac{MU-MIMO} scenario where $K$ single-antenna \acp{UE} communicate with an $M$-antenna gigantic \ac{MIMO} \ac{BS}, with $M\gg K$. The $M\times 1$ received complex \ac{BB} vector can be expressed as
\begin{equation}
\boldsymbol{y} = \boldsymbol{H}\boldsymbol{s} + \boldsymbol{n},
\label{eq:ul_model}
\end{equation}
where $\boldsymbol{H}$ is the $M\times K$ channel matrix, $\boldsymbol{s}$ is the $K \times 1$ vector of symbols transmitted by the users, and $\boldsymbol{n}\sim \mathcal{CN}(\bs{0},N_0 \mathbf{I}_{M})$ is the additive white Gaussian noise (AWGN) vector. 

\begin{figure}[h]
    \centering
    \includegraphics[scale=0.5]{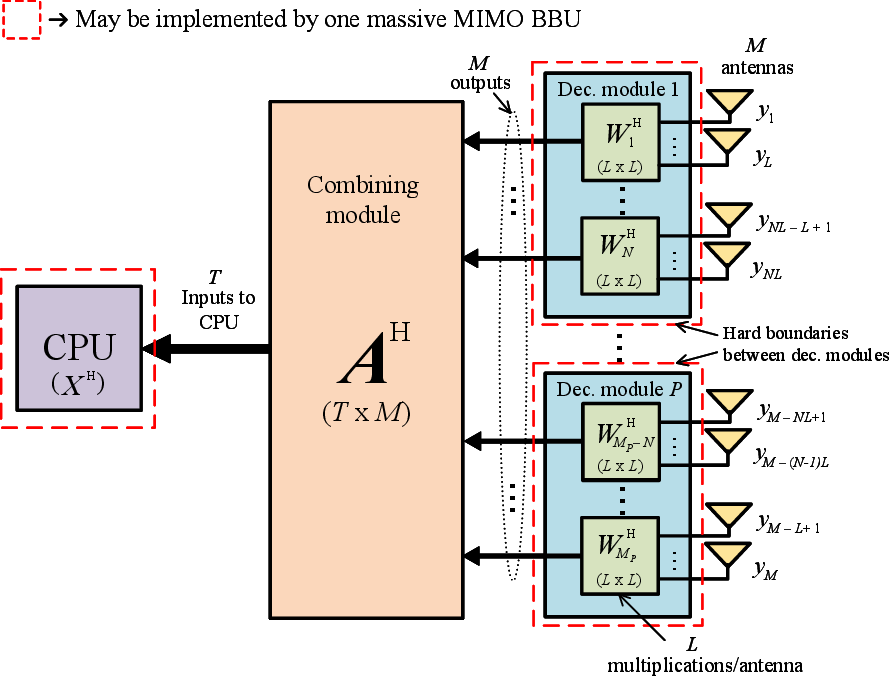}
    \caption{Schematic of the system \ac{UL} data processing.}\label{fig:syst_mod}
\end{figure}

The \ac{BB} processing is performed based on the WAX framework \cite{wax_journal,A_struct_journal,cell-free_wax}, which allows for an arbitrary trade-off between level of decentralization (inputs to \ac{CPU} $T$) and decentralized processing complexity (multiplications per antenna $L$). Similar to \cite{cell-free_wax}, we consider that the $\MP\triangleq M/L$ decentralized linear filters of dimension $L\times L$ are distributed among $P$ decentralized modules. We assume that each decentralized module independently computes and applies an equal (integer) number $N=\MP/P$ of decentralized filters. A schematic of the system during \ac{UL} data processing is shown in Fig.~\ref{fig:syst_mod}. The resulting post-processed vector may be expressed as
\begin{equation}\label{eq:post_proc_vec}
    \bs{z}=\bs{X}^\mrm{H}\bs{A}^\mrm{H}\bs{W}^\mrm{H}\bs{y},
\end{equation}
where $\bs{W}=\diag(\bs{W}_{1},\dots,\bs{W}_{\MP})$ is an $M\times M$ block diagonal matrix, with the $L\times L$ diagonal blocks $\{\bs{W}_{m}\}_{m=1}^{\MP}$ corresponding to the decentralized filters, $\bs{A}$ is an $M\times T$ matrix, associated to the combining module reducing the inputs to the \ac{CPU}, and $\bs{X}$ is a $T\times K$ matrix, associated to the linear equalizer potentially applied at the \ac{CPU}. We assume that the $m$th decentralized filter $\bs{W}_{m}$ is applied at decentralized module $p=1+\big\lfloor\frac{m-1}{N}\big\rfloor$, so that each decentralized module applies $N$ contiguous decentralized filters.

Unlike in previous work considering the WAX framework \cite{wax_journal,A_struct_journal, cell-free_wax}, we hereby impose the restriction that the different decentralized modules are not allowed to cooperate with each other for computing their decentralized filters, i.e., they are separated by hard boundaries. Note that, although \cite{cell-free_wax} considers a system model identical to the one in Fig.~\ref{fig:syst_mod} during the \ac{UL} data phase, no discussion was made in that work about the computation of the decentralized filters, which requires some level of cooperation among decentralized modules (and/or \ac{CPU}) even when considering the available decentralized approaches for the WAX framework \cite{A_struct_journal}. If we restrict $\TP \triangleq T/L$ to be an integer, we may also consider the following practical restriction for the combining module proposed in \cite{A_struct_journal}:
\begin{equation}\label{eq:rest_comb_kron}
    \bs{A}=\widetilde{\bs{A}}\otimes \mathbf{I}_L,
\end{equation} 
where $(\cdot)\otimes(\cdot)$ denotes the Kronecker product, $\widetilde{\bs{A}}=\big[\widetilde{\bs{A}}_\mrm{T}^\mrm{T},\widetilde{\bs{A}}_\mrm{B}^\mrm{T}\big]^\mrm{T}$ is an $\MP \times \TP$ matrix, which may be divided into its $\TP \times \TP$ upper block $\widetilde{\bs{A}}_\mrm{T}$ and its $(\MP-\TP) \times \TP$ lower block $\widetilde{\bs{A}}_\mrm{B}$. The structure in \eqref{eq:rest_comb_kron} is associated with a number of practical benefits described in \cite{A_struct_journal}, while it facilitates the design of specific combining modules that enable decentralized schemes to compute the decentralized filters.

\subsection{Preliminaries on the WAX framework}
In \cite{wax_journal,A_struct_journal}, important theoretical results were derived for the WAX framework. We will summarize the main results upon which we will base our particularization to gigantic \ac{MIMO} systems with hard boundaries.

In \cite{wax_journal}, the achievability of information-lossless processing in the WAX framework was shown to be equivalent to the ability to decompose the channel matrix $\bs{H}$ into the so-called WAX decomposition, given by \cite[Theorem~1]{wax_journal}
\begin{equation}\label{eq:WAX}
    \bs{H}=\bs{W}\bs{A}\bs{X},
\end{equation}
where the matrices have direct correspondence with the processing applied in \eqref{eq:post_proc_vec}, making it directly applicable to the system model described above. This decomposition could be thus exploited to characterize the information lossless trade-off between $L$ and $T$, given in \cite[Theorem~1]{wax_journal} as
\begin{equation}\label{eq:wax_to}
    L>K\frac{M-T}{M} = K\frac{\MP-\TP}{\MP},
\end{equation}
which assumes an unstrucutred $\bs{A}$, e.g., a realization of an i.i.d. Gaussian matrix, and where we have ignored the regime where $T\geq K$ since it trivially leads to information loss.

Taking into account the structure in \eqref{eq:rest_comb_kron}, with the extra restriction of having full-rank $\tA_{\mrm{T}}$, \cite[Corollary~1]{A_struct_journal} provides an alternative formulation of the WAX decomposition where $\bs{X}$ is conveniently substituted. In such formulation, information-lossless processing can be attained by finding the decentralized filters $\{\bs{W}_{m}\}_{m=1}^{\MP}$ solving
\begin{equation}\label{eq:WAX_face}
    \begin{bmatrix}\W_{1}^{-1} & \dots & \W_{\MP}^{-1}\end{bmatrix} \left( \Btil \bullet \H \right)=\boldsymbol{0}_{L\times K(\MP-\TP)},
\end{equation}
where $\Btil \bullet \H$ is the face-splitting product between $\Btil$ and $\bs{H}$, whose rows are given by the Kronecker product of the rows of $\Btil$ with the $L\times K$ row-blocks of $\bs{H}=[\bs{H}_{1}^\mrm{T} \;\dots \; \bs{H}_{\MP}^\mrm{T}]^\mrm{T}$, and where we have
\begin{equation}\label{eq:B_def}
\Btil = \begin{bmatrix} \widetilde{\A}_\text{B}\widetilde{\A}_\text{T}^{-1} & -\mathbf{I}_{\MP-\TP}\end{bmatrix}^\mathrm{T}.
\end{equation}
The matrix $\Btil$ captures the applicability of the respective combining module $\bs{A}$, as well as the restrictions on the available decentralized schemes to compute the decentralized filters $\{\bs{W}_{m}\}_{m=1}^{\MP}$. 

The alternative formulation of the WAX decomposition from \eqref{eq:WAX_face} was exploited in \cite{A_struct_journal} to define sparse constructions for $\bs{A}$ based on sum modules. These constructions led to an achievable trade-off, corresponding to a degenerate version of \eqref{eq:wax_to} when fixing one of the decentralized filters, given by
\begin{equation}\label{eq:wax_ach_to}
    L\geq K\frac{\MP-\TP}{\MP-1}.
\end{equation}

\subsection{Problem formulation}
 The main goal of this work is to define a suitable combining module $\bs{A}$, or equivalently $\Btil$, such that each decentralized module can find the decentralized filters that it should apply without cooperating with other decentralized modules. Specifically, decentralized module $p\in \{1,\dots,P\}$ should be able to independently find the set of filters $\{\bs{W}_{m}\}_{m\in \mathcal{P}_p}$ that give a solution to \eqref{eq:WAX}, equivalently to \eqref{eq:WAX_face}, where
\begin{equation}\label{eq:group_indexing}
    \mathcal{P}_p = \{m\in \mathbb{N}: (p-1)N+1\leq m \leq p N)\}
\end{equation}
corresponds to the set of indices of the decentralized filters applied at decentralized module $p$. We further consider that each decentralized module is in charge of estimating the channel matrix associated to its own antennas. Hence, decentralized module $p\in\{1,\dots,P\}$ only has access to the estimates of the channel blocks $\{\bs{H}_{m}\}_{m\in \mathcal{P}_p}$, where we focus on perfect channel estimation for ease of presentation.

\subsection{Applicability to Gigantic MIMO}
The considered system model is especially suitable to gigantic \ac{MIMO} scenarios. The reason is that, to facilitate backward compatibility and ease of deployment, this technology may reuse some of the hardware modules already implemented in massive \ac{MIMO} \acp{BS}, as the \acp{BBU}. Considering an $F\times$ increase in the number of antennas of gigantic \ac{MIMO}, the \ac{BB} processing may then need to be divided into $P=F$ independent decentralized modules, each corresponding to a massive \ac{MIMO} \ac{BBU} connected to $M/F$ antennas as in the original massive \ac{MIMO} setting. Moreover, since this framework allows for an arbitrary number of inputs to the \ac{CPU}, we could select $T=M/F$ so that the \ac{CPU} may be implemented by another massive \ac{MIMO} \ac{BBU}, i.e., without having to redesign the \acp{BBU} to support the increased number of inputs. The previous correspondence is further indicated in Fig.~\ref{fig:syst_mod}. Note that the considered framework is also applicable to \ac{DL}, as explained in \cite{wax_journal}.

In the unconstrained case (no hard-boundaries), the choice $T=M/F$ allows for information-lossless processing by increasing $L$ sufficiently, as seen from \eqref{eq:wax_to} and \eqref{eq:wax_ach_to}. The only necessary condition is to have $M/F\geq K$, which is a reasonable assumption in gigantic \ac{MIMO} scenarios where $M\gg K$. In this work, we will show that this property is also applicable in the presence of hard-boundary constraints, hence enabling the implementation of gigantic \ac{MIMO} systems from state-of-the-art massive \ac{MIMO} \acp{BBU}.

\section{Combining modules with hard boundaries}
The alternative formulation of the WAX decomposition \eqref{eq:WAX_face} reveals in the columns of $\Btil$ which decentralized modules should interact with each other to be able to attain a solution. Specifically, the non-zero elements of each column of $\Btil$ indicate the $\bs{W}_{m}$ blocks to be included in the resulting sum, after multiplication with the respective block of $\bs{H}$. Given the structure of $\Btil$ in \eqref{eq:B_def}, we can note that, even if $\tA_{B}$ is conveniently sparsified, the decentralized filters $\bs{W}_{m}$ interacting with each other in \eqref{eq:WAX_face} may not be restricted to a single decentralized module. However, an arbitrary re-indexing of $\{\bs{W}_{m}\}_{m=1}^{\MP}$ gives a permutation that may be absorbed in the resulting $\tA$ without affecting the solvability of \eqref{eq:WAX} \cite[Proposition~2]{A_struct_journal}. We thus consider a suitable re-indexing, given by the permutation function $\pi(\cdot)$, for which we define
\begin{equation}
\begin{aligned}
    \bs{W}^{\pi} &\triangleq \diag(\bs{W}_{\pi(1)},\dots,\bs{W}_{\pi(\MP)})\\
    &=(\bs{P}^{\pi}\otimes \mathbf{I}_L)\bs{W} (\bs{P}^{\pi}\otimes \mathbf{I}_L)^\mrm{T},
\end{aligned}
\end{equation}
where $\bs{P}^{\pi}$ is the $\MP\times\MP$ permutation matrix associated to the permutation function $\pi(\cdot)$. To achieve information-lossless processing, we may then consider the permuted version of the WAX decomposition \eqref{eq:WAX} as
\begin{equation}\label{eq:WAX_perm}
    \bs{W}^{\pi} \bs{A}^{\pi} \bs{X} = \bs{H}^{\pi},
\end{equation}
where $\bs{H}^{\pi}\triangleq(\bs{P}^{\pi}\otimes \mathbf{I}_L)\bs{H}=[\bs{H}_{\pi(1)}^\mrm{T}, \dots, \bs{H}_{\pi(\MP)}^\mrm{T}]^\mrm{T}$, and 
\begin{equation}\label{eq:A_perm}
    \bs{A}^{\pi}\triangleq(\bs{P}^{\pi}\otimes \mathbf{I}_L)\bs{A} = (\tA^{\pi} \otimes \mathbf{I}_{L}),  
\end{equation}
with $\tA^{\pi}=\bs{P}^{\pi}\tA$. Note that \eqref{eq:WAX_perm} has full correspondence with \eqref{eq:WAX} by left multiplying both sides with $(\bs{P}^{\pi}\otimes \mathbf{I}_L)$. Hence, we can obtain a solution to \eqref{eq:WAX} by solving \eqref{eq:WAX_perm} with a valid combining module $\bs{A}^{\pi}$, and permuting back the combining module and resulting decentralized filters. We next focus on designing a valid combining module to solve \eqref{eq:WAX_perm} such that the solution only requires interaction within $P$ mutually-exclusive sets of decentralized filters. Subsequently, we will characterize suitable permutations $\pi(\cdot)$ such that the non-intersecting groups coincide with those indexed by \eqref{eq:group_indexing} after permuting back $\bs{W}$.

\subsection{Solving the WAX decomposition via mutually-exclusive decentralized filter sets}
Taking \eqref{eq:A_perm} into account, and considering the division $\tA^{\pi}=[(\tA^{\pi}_\mrm{T})^\mrm{T}, (\tA^{\pi}_\mrm{B})^\mrm{T}]^\mrm{T}$ into the top $\TP\times \TP$ and bottom $(\MP-\TP)\times \TP$ blocks, we may now apply the alternative formulation of the WAX decomposition to \eqref{eq:WAX_perm}. This gives
\begin{equation}\label{eq:WAX_face_perm}
    \begin{bmatrix}\W_{\pi(1)}^{-1} & \dots & \W_{\pi(\MP)}^{-1}\end{bmatrix} \left( \Btil^{\pi} \bullet \H^{\pi} \right)=\boldsymbol{0}_{L\times K(\MP-\TP)},
\end{equation}
where $
\Btil^{\pi} = \begin{bmatrix} \widetilde{\A}_\text{B}^{\pi}(\widetilde{\A}_\text{T}^{\pi})^{-1} & -\mathbf{I}_{\MP-\TP}\end{bmatrix}^\mathrm{T}$. Note that, to be able to apply the alternative WAX formulation to \eqref{eq:WAX_perm}, we should now have full-rank $\widetilde{\A}_\text{T}^{\pi}$, corresponding to the permuted version. Considering \cite[Proposition~1]{A_struct_journal}, we have full freedom in selecting said matrix, so we will assume without loss of generality $\widetilde{\A}_\text{T}^{\pi}=\mathbf{I}_{\TP}$, which further leads to a desirable sparse structure. The interaction among different decentralized filters is determined by the indices of the non-zero entries at different columns of $(\tA^{\pi}_\mrm{B})^\mrm{T}$ (or rows of $\tA^{\pi}_\mrm{B}$), which will still inevitably require interaction with the decentralized filters associated to the non-zero entries from the respective columns of the $-\mathbf{I}_{\MP-\TP}$ block in $\Btil^{\pi}$. In order to have $P$ non-intersecting sets, we should then enforce a block diagonal structure on $\tA_{B}$ with $P$ diagonal blocks, i.e.,
\begin{equation}\label{eq:Ab_diag}
\tA_\mrm{B}^{\pi}=\diag (\tA_{\mrm{B},1}^{\pi},\dots,\tA_{\mrm{B},P}^{\pi}),
\end{equation}
where block $\tA_{\mrm{B},p}^{\pi}$ corresponds to an $\frac{\MP-\TP}{P} \times \frac{\TP}{P}$ matrix. Note the extra requirement of $\TP$ being a multiple of $P$. Substituting \eqref{eq:Ab_diag} in \eqref{eq:WAX_face_perm} leads to a set of $P$ independent equations, with equation $p$ given in \eqref{eq:p_WAX_indep_eq} at the end of this page. Since these equations include non-intersecting sets of decentralized filters, they can be solved in parallel at different decentralized modules, ensuring the hard-boundary constraints. 

\begin{rmk}
We may note that the indices of the channel submatrices in \eqref{eq:p_WAX_indep_eq} coincide with those of the decentralized filters. This means that, assuming that the permutation function $\pi(\cdot)$ is suitably designed (as outlined in the next subsection) each decentralized module only requires knowledge of the local channels $\{\bs{H}_{\pi(m)}\}_{\pi(m)\in \mathcal{P}_p}$ available at that module. Thus, no sharing of information is required among the decentralized modules in order to solve \eqref{eq:p_WAX_indep_eq}.
\end{rmk}

\begin{figure*}[b]
\vspace{-1em}
\hrulefill
\begin{equation}\label{eq:p_WAX_indep_eq}
\begin{aligned}
    \Big[\bs{W}^{-1}_{\pi\left((p-1)\frac{\TP}{P}+1\right)} \; \cdots \; \bs{W}^{-1}_{\pi\left(p\frac{\TP}{P}\right)} \; \; \bs{W}^{-1}_{\pi\left(\TP+(p-1)\frac{\MP-\TP}{P}+1\right)} &\;  \cdots \; \bs{W}^{-1}_{\pi\left(\TP+p\frac{\MP-\TP}{P}\right)}\Big]\\
    \times & \left( \begin{bmatrix}
        \big(\tA^{\pi}_{B,p}\big)^\mrm{T} \\
        -\mathbf{I}_{\frac{\MP-\TP}{P}}
    \end{bmatrix}\bullet\begin{bmatrix}
        \bs{H}_{\pi\left(1+(p-1)\frac{\TP}{P}\right)} \\
        \vdots \\
        \bs{H}_{\pi\left(p\frac{\TP}{P}\right)}\\
        \bs{H}_{\pi\left(\TP+(p-1)\frac{\MP-\TP}{P}+1\right)}\\
        \vdots\\
        \bs{H}_{\pi\left(\TP+p\frac{\MP-\TP}{P}\right)}
    \end{bmatrix}\right)=\bs{0}_{L\times K\frac{\MP-\TP}{P}}
\end{aligned}
\end{equation}

\vspace{-1em}
\end{figure*}

Equation \eqref{eq:p_WAX_indep_eq}, which may be solved independently at each decentralized module, has full correspondence with the alternative WAX formulation \eqref{eq:WAX_face} for a system where $\MP$ and $\TP$ are substituted by (integer) $\MP/P$ and $\TP/P$, respectively. All the results from \cite{A_struct_journal} on its solvability, as well as on the validity of $\tA_{\mrm{B},p}$ matrices, are thus directly applicable to this case by simply performing such substitution. This leads to an achievable trade-off given by
\begin{equation}\label{eq:ach_new_tradeoff}
    L\geq K\frac{\MP-\TP}{\MP-P},
\end{equation}
which is a degenerate version of \eqref{eq:wax_ach_to} under structured $\bs{A}$, assuming now that one decentralized filter per decentralized module is kept fixed so that the techniques from \cite{A_struct_journal} are directly applicable within each decentralized module.

 If a solution to \eqref{eq:p_WAX_indep_eq} exists, we can find it by vectorizing the equation and selecting the resulting vector $\mrm{vec}([\bs{W}^{-1}_{\pi(m)}]_{\pi(m)\in\mathcal{P}_p})$ from the null-space of the factor matrix obtained from \eqref{eq:p_WAX_indep_eq} after vectorization. On the other hand, assuming that we have a valid structure for $\bs{A}$, \cite[Theorem~2]{wax_journal} ensures that a randomly chosen point from said null-space will lead to full-rank $\bs{W}^{-1}_{m}$ matrices with probability 1 (after inverse vectorization) so that the respective $\bs{W}_m$ filters can be applied to attain information-lossless processing. In case information-lossless processing is not available, we may consider an approximate least squares solution to the vectorized version of \eqref{eq:p_WAX_indep_eq}, solved by selecting $\mrm{vec}([\W_{\pi(m)}^{-1}]_{\pi(m)\in \mathcal{P}_{p}})$ as the eigenvector associated to the lowest eigenvalue of the factor matrix. Note that both approaches coincide in the information-lossless regime under valid $\bs{A}$. However, for the information-lossy regime, we cannot ensure that the resulting solution attains full-rank $\bs{W}^{-1}_{m}$. This problem can be fixed by randomly combining $N_{\mrm{eig}}\geq 1$ eigenvectors associated to the $N_{\mrm{eig}}$ lowest eigenvalues, and iteratively increasing $N_{\mrm{eig}}$ until the full-rank requirement is achieved.

\subsection{Connecting mutually-exclusive decentralized filters sets with decentralized modules}
We now focus on finding a permutation $\pi(\cdot)$ ensuring that the decentralized filters obtained at each decentralized module $p$ coincide with the ones applied at said module during the data phase. This corresponds to having the subindices in each non-interfering equation \eqref{eq:p_WAX_indep_eq} contained within the set $\mathcal{P}_p$ defined in \eqref{eq:group_indexing}. We may thus select any permutation $\pi(\cdot)$ that fulfills 
\begin{subequations}\label{eq:cons_perm}
\begin{equation}
    \pi\left((p-1)\frac{\TP}{P}+i\right)\in \mathcal{P}_p,
\end{equation}
\begin{equation}
    \pi\left(\TP+(p-1)\frac{\MP-\TP}{P}+j\right) \in \mathcal{P}_p,
\end{equation}
\end{subequations}
$\forall p \in \{1,\dots, P\}$, $\forall i \in \left\{1,\dots, \frac{\TP}{P}\right\}$, and $\forall j=\left\{1,\dots,\frac{\MP-\TP}{P}\right\}$. Since $\mathcal{P}_p$, $\forall p$, are mutually exclusive sets, each composed of $N$ elements, while \eqref{eq:cons_perm} consists of $N=\frac{\MP}{P}$ permutation entries for each of the $P$ mutually exclusive sets indexed by $p$ in \eqref{eq:p_WAX_indep_eq}, we can always find a suitable $\pi(\cdot)$. In fact, we have a total of $(N!)^P$ permutation functions $\pi(\cdot)$ fulfilling \eqref{eq:cons_perm}. One example is simply defining said permutation as
\begin{subequations}
\begin{equation}
    \pi\left((p-1)\frac{\TP}{P}+i\right) = (p-1)N+i,
\end{equation}
\begin{equation}
    \pi\left(\TP+(p-1)\frac{\MP-\TP}{P}+j\right) = (p-1)N+\frac{\TP}{P}+j,
\end{equation}  
\end{subequations}
$\forall p \in \{1,\dots, P\}$, $\forall i \in \left\{1,\dots, \frac{\TP}{P}\right\}$, and $\forall j=\left\{1,\dots,\frac{\MP-\TP}{P}\right\}$.
\section{Numerical results}
In the previous section, we showed how to design the combining module in the WAX framework such that the decentralized processing may be independently characterized and applied by a set of decentralized modules with hard-boundaries between them. We will now numerically analyze the potential of this approach in a gigantic \ac{MIMO} scenario.

We consider a \ac{BS} with $M=256$ antennas, which are divided into $4$ decentralized modules, each of which may be implemented using a single massive MIMO \ac{BBU} designed to handle $N=64$ antennas. The decentralized modules are connected to a fixed combining module, which is constructed as described in the previous sections, where, for simplicity, $\tA_{\mrm{B},p}$ has been randomly chosen $\forall p$ by taking one realization of Gaussian random matrix. The combining module generates $T=64$ inputs to the \ac{CPU}, such that it can also be implemented using a single massive MIMO \ac{BBU}. 

\begin{figure}[h]
    \centering
    \begin{subfigure}[h]{0.49\textwidth}
         \centering         
         \includegraphics[scale=0.48]{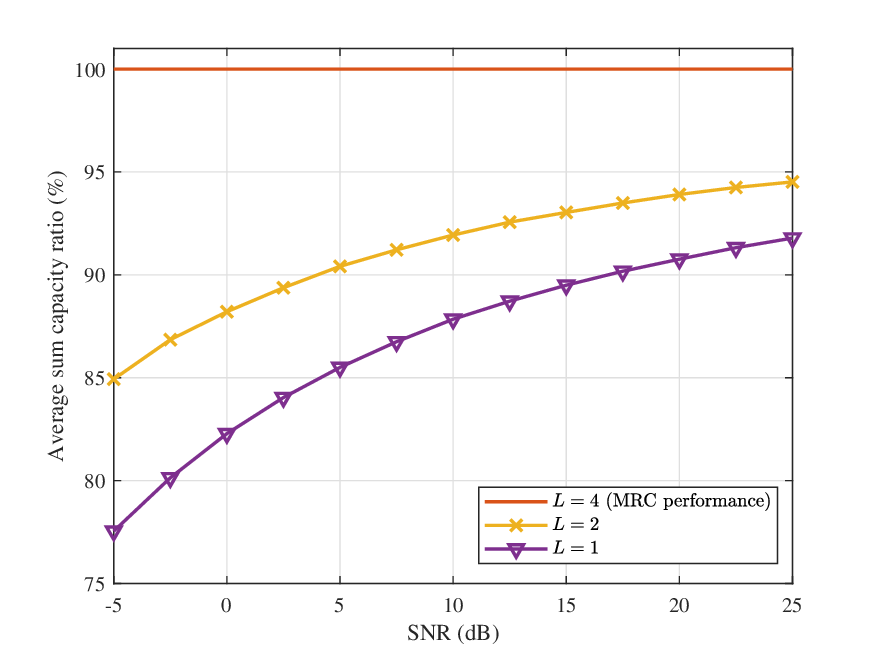}
         \caption{$K=5$}
         \label{fig:K_5}

     \end{subfigure}
     \begin{subfigure}[h]{0.49\textwidth}
         \centering
    \includegraphics[scale=0.48]{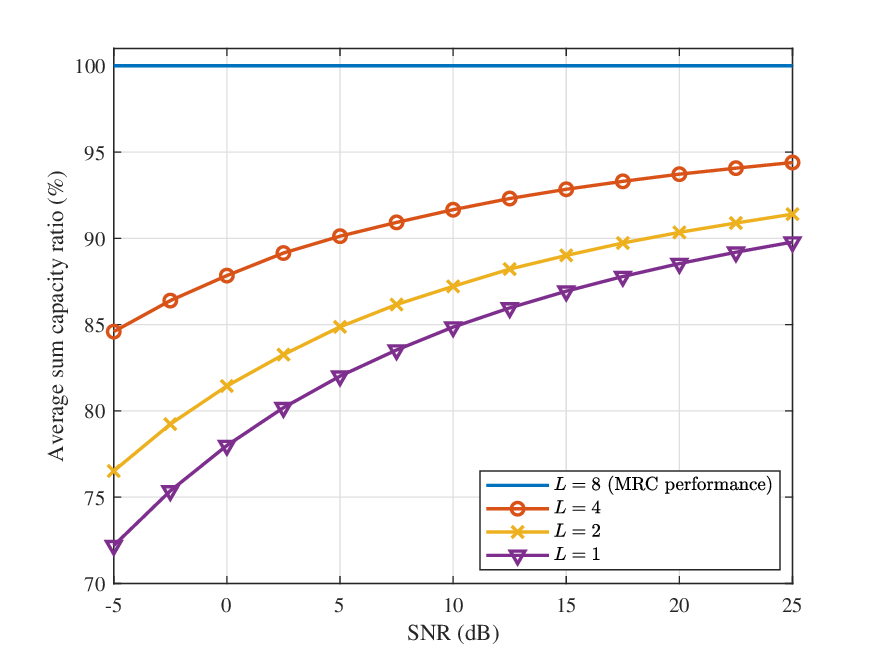}
    \caption{$K=9$}
    \label{fig:K_9}
     \end{subfigure}

    \caption{Average sum capacity ratio with respect to \ac{SNR}.}
    \label{fig:rates}
    \vspace{-2em}
\end{figure}

Fig.~\ref{fig:rates} shows the average sum capacity ratio, corresponding to the ratio between the channel capacity for \eqref{eq:ul_model} and the capacity after the decentralized processing. We have averaged throughout $10^3$ realizations of a standard i.i.d Rayleigh fading channel, but the channel model should not impact the results significantly as long as it classifies as randomly chosen according to \cite{wax_journal,A_struct_journal}. For comparison, we may take the simplest information-lossless processing, namely \ac{MRC}, which may be implemented in a decentralized fashion requiring $K$ multiplications per antenna and $K$ inputs to the \ac{CPU} \cite{wax_journal}. Note that this approach inevitably wastes some of the available \ac{CPU} inputs when it is implemented through a massive \ac{MIMO} \ac{BBU}. The numerical results show that the proposed approach can attain the information-lossless \ac{MRC} performance with $L=8$ multiplications per antenna for $K=5$ in Fig.~\ref{fig:K_5}, and with $L=4$ multiplications per antenna for $K=9$ in Fig.~\ref{fig:K_9}, both of which tightly fulfill \eqref{eq:ach_new_tradeoff}. This still allows for a reduction in the decentralized complexity compared to \ac{MRC}, as seen in Table~\ref{table:Complx}. Moreover, the proposed framework allows for further complexity reductions at the cost of minor performance loss, since for reasonable \ac{SNR} values can achieve $90\%$ of the capacity with half the decentralized processing complexity.

\begin{table}[h]
\caption{Processing complexity at decentralized modules.}
\centering
\begin{tabularx}{0.46\textwidth}{ |c| *{3}{Y|}}
\cline{2-3}
\multicolumn{1}{c|}{} &  \multicolumn{2}{c|}{MAC operations per dec. module} \\
\cline{2-3}
\multicolumn{1}{c|}{} & $K=9$ & $K=5$ \\
\cline{1-3}
 MRC baseline & 576 & 320 \\    
 \cline{1-3}
 $L=8$ & 512 & - \\  
 \cline{1-3}
 $L=4$ & 256 & 256\\  
 \cline{1-3}
 $L=2$ & 128 & 128 \\  
 \cline{1-3}
 $L=1$ & 64 & 64 \\  
 \cline{1-3}
\end{tabularx}
\label{table:Complx}
\end{table}

\section{Conclusion}
We have studied an adaptation of the WAX framework to facilitate the implementation of gigantic \ac{MIMO}. The proposed approach allows applying information-lossless decentralized processing at non-cooperating decentralized modules, such that they may be implemented using state-of-the-art massive \ac{MIMO} \acp{BBU}. We have also shown how this framework may be exploited to reduce decentralized processing complexity at the cost of minor performance loss.
\bibliographystyle{IEEEtran}
\bibliography{IEEEabrv,wax}

%
\IEEEpeerreviewmaketitle

\end{document}